\documentclass[twocolumn,showpacs,amsmath,amssymb,prl,superscriptaddress]{revtex4-1}
\usepackage{graphicx}
\usepackage{dcolumn}
\usepackage{bm}
\usepackage {longtable}
\usepackage{braket}

\usepackage{color}      
\newcommand \be{\begin{equation}}
\newcommand \ee{\end{equation}}
\newcommand \bea{\begin{eqnarray}}
\newcommand \eea{\end{eqnarray}}
\newcommand \bse{\begin{subequations}}
\newcommand \ese{\end{subequations}}

\begin{document}


\title{Quantum gates in mesoscopic atomic ensembles based on adiabatic passage and Rydberg blockade}

\author{I.~I.~Beterov}
\email{beterov@isp.nsc.ru}
\affiliation{A.V.Rzhanov Institute of Semiconductor Physics SB RAS, 630090 Novosibirsk, Russia}
\affiliation{Novosibirsk State University, 630090 Novosibirsk, Russia}

\author{M.~Saffman}
\affiliation{Department of Physics, University of Wisconsin, Madison, Wisconsin, 53706, USA}

\author{E.~A.~Yakshina}
\affiliation{A.V.Rzhanov Institute of Semiconductor Physics SB RAS, 630090 Novosibirsk, Russia}
\affiliation{Novosibirsk State University, 630090 Novosibirsk, Russia}

\author{V.~P.~Zhukov}
\affiliation{Institute of Computational Technologies SB RAS, 630090 Novosibirsk, Russia}

\author{D.~B.~Tretyakov}
\affiliation{A.V.Rzhanov Institute of Semiconductor Physics SB RAS, 630090 Novosibirsk, Russia}

\author{V.~M.~Entin}
\affiliation{A.V.Rzhanov Institute of Semiconductor Physics SB RAS, 630090 Novosibirsk, Russia}

\author{I.~I.~Ryabtsev}
\affiliation{A.V.Rzhanov Institute of Semiconductor Physics SB RAS, 630090 Novosibirsk, Russia}
\affiliation{Novosibirsk State University, 630090 Novosibirsk, Russia}
\affiliation{Russian Quantum Center, Skolkovo, Moscow Reg., 143025, Russia}

\author{C.~W.~Mansell}
\affiliation{The Open University, Walton Hall, MK7 6AA, Milton Keynes, UK}

\author{C.~MacCormick}
\affiliation{The Open University, Walton Hall, MK7 6AA, Milton Keynes, UK}

\author{S.~Bergamini}
\affiliation{The Open University, Walton Hall, MK7 6AA, Milton Keynes, UK}

\author{M.~P.~Fedoruk}
\affiliation{Novosibirsk State University, 630090 Novosibirsk, Russia}
\affiliation{Institute of Computational Technologies SB RAS, 630090 Novosibirsk, Russia}

\date{16 April 2012}

\begin{abstract}
We present schemes for geometric phase compensation in adiabatic passage which can be used for the implementation of
quantum logic gates with atomic ensembles consisting of an arbitrary number of strongly interacting atoms. Protocols
using double sequences of stimulated Raman adiabatic passage (STIRAP) or adiabatic rapid passage (ARP) pulses are
analyzed. Switching the sign of the detuning between two STIRAP sequences, or inverting the phase between two ARP
pulses, provides state transfer with well defined amplitude and phase independent of atom number in the Rydberg
blockade regime. Using these pulse sequences we present protocols for universal single-qubit and two-qubit operations
in atomic ensembles containing an unknown number of atoms.
\end{abstract}

\maketitle

Quantum information can be stored in  collective states of  ensembles of strongly interacting atoms~\cite{Lukin2001}.  This idea can  be extended to encoding an entire register of qubits in ensembles of atoms with multiple ground states~\cite{Brion2007d} which opens up the possibility of large quantum registers in a single atomic ensemble~\cite{Saffman2008}, or of coupling arrays of small ensembles in a scalable atom chip based architecture~\cite{Whitlock2009}. Quantum information based on ensembles can be realized  more generally in any  ensemble of strongly coupled spins~\cite{Tordrup2008,*Wesenberg2009}. Our proposal for implementing high fidelity quantum gates in ensembles is thus of interest for several different implementations of quantum computing.

The enhanced coupling to the radiation field by a factor of $\sqrt N$, with $N$ the number of atoms or spins, is useful
for coupling matter qubits to single photons~\cite{Dudin2012,*Peyronel2012}. Combining photon coupling with local
quantum gates in ensembles enables architectures with improved fidelity for quantum
networking~\cite{Jiang2007,*Pedersen2009}.  The use of ensemble qubits is also attractive  for deterministic loading of
registers of single atom qubits~\cite{Saffman2002,Muller2009,*Saffman2009b,Isenhower2011} and for realizing gates that
act on multiple particles. All of these capabilities rely on high fidelity  quantum gate operations between
collectively encoded qubits. However, due to the dependence of the Rabi frequency of oscillations between different
collective states on the number of atoms as  $\sqrt N$,  it is difficult to perform gates with well defined rotation
angles in the situation where $N$ is unknown~\cite{Saffman2010, Whitlock2010b}. Although there is recent progress in
nondestructive  measurement of  $N$ with high accuracy~\cite{HZhang2012} it remains an outstanding challenge to
implement high fidelity quantum logic gates without precise knowledge of $N$, particularly in the case of collectively
encoded registers~\cite{Brion2007d} where the effective value of $N$ depends on the unknown quantum state encountered
during a computation.

\begin{figure}[!t]
\includegraphics[width=\columnwidth]{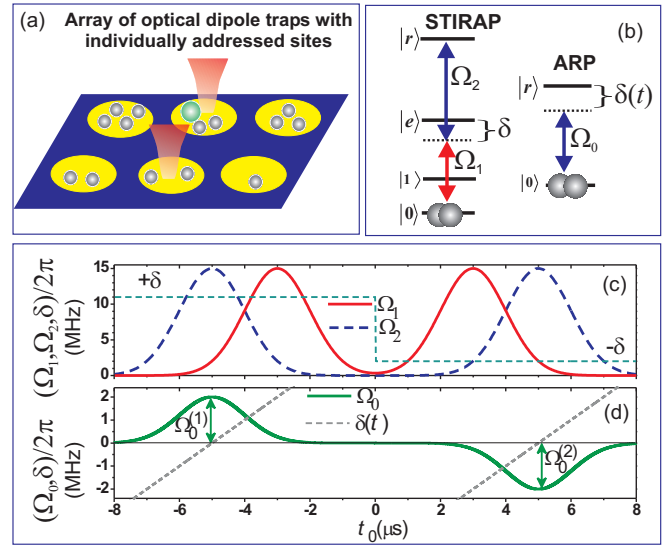}
\vspace{-.5cm}
\caption{
\label{LevelScheme}(Color online).
(a) Scheme of the quantum register based on individually addressed atomic ensembles in the array of optical dipole traps. Laser pulses are used to excite atoms into the Rydberg state. Only one atom in each site can be excited due to Rydberg blockade. Simultaneous excitation of Rydberg atoms in the neighboring sites is also blocked; (b) Energy levels for two-photon STIRAP and single-photon ARP excitation; (c) Time sequence of
STIRAP laser pulses; (d) Time sequence for ARP laser excitation;
}
\end{figure}

Adiabatic passage techniques (STIRAP and ARP) have been widely used  for deterministic population transfer in atomic
and molecular systems~\cite{Bergmann1998,Abragam1961}. These techniques have been studied for  quantum state
control~\cite{Marte1991}, qubit rotations~\cite{Kis2002},  creation of entangled states~\cite{Moller2008}, and for
deterministic excitation of Rydberg atoms~\cite{Pohl2010,Beterov2011}. Although STIRAP or ARP methods provide  pulse
areas with strongly suppressed sensitivity to  the Rabi frequency $\Omega_N$, and therefore suppressed sensitivity to
$N$, the phase of the final state is in general still strongly dependent on $N$. Randomly loaded dipole trap follow a
Poissonian distribution in the atom number, with relative fluctuations $1/ \sqrt{N}$. Indeed gate errors at the level
of $10^{-3}$ can be achieved, but would require $\bar N\sim 4000$, and achieving full blockade for such a large
ensemble remains an outstanding challenge.

In this Letter we propose double adiabatic sequences using either STIRAP or ARP excitation  which remove the phase
sensitivity, and can be used to implement gates on collectively encoded qubits without precise knowledge of $N$ even
for moderate size ensembles.

\textit{Method for phase compensation:} Our approach is shown in Fig.~\ref{LevelScheme}. The quantum register consists
of individually addressed atomic ensembles in  arrays of optical dipole traps or  optical lattices
[Fig.~\ref{LevelScheme}(a)]. The  energy levels scheme for STIRAP and ARP is shown on Fig.~\ref{LevelScheme}(b).
 A sequence of two STIRAP pulses is produced with fields having  Rabi frequencies $\Omega _{1}$, $ \Omega _{2}$, and
detuning $\delta $ from the intermediate state.  In the regime of strong Rydberg blockade, the first STIRAP (ARP) pulse
deterministically prepares the ensemble in a collective state with a single Rydberg excitation, as we demonstreted in
\cite{Beterov2011}. The second reverse STIRAP pulse, as shown in Fig.~\ref{LevelScheme}(c), returns the Rydberg atom
back to the ground state. Similar scheme can be implemented using linearly chirped ARP pulses, as shown in
Fig.~\ref{LevelScheme}(d).

\begin{figure}[!t]
\includegraphics[width=\columnwidth]{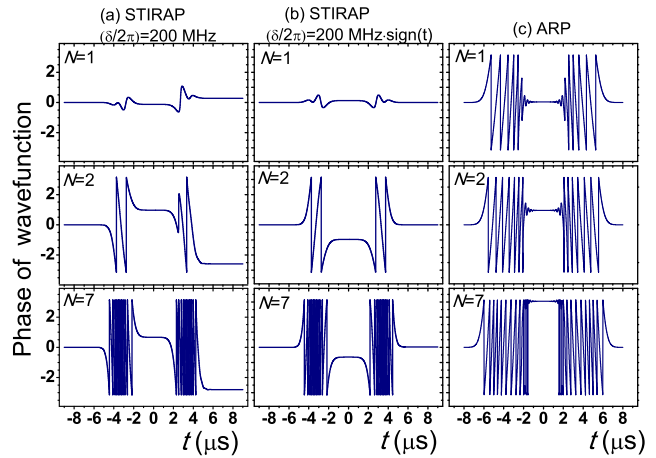}
\vspace{-.6cm}
\caption{
\label{Phases}(Color online).
Calculated time dependence of the phase of the collective ground state amplitude for $N=1,2,7$ atoms (top to bottom).
Double STIRAP sequence [$\Omega_1/2\pi = 30~\rm MHz,$ $\Omega_2/2\pi = 40~\rm MHz$]   with  $\delta/2\pi = 200~\mathrm{MHz}$ (a), with  $\delta/2\pi = 200~\mathrm{MHz} \times \mathrm{sign}\left( t \right)$ (b), and for a double ARP pulse sequence with phase inversion (c). The single STIRAP sequence used $\Omega_{j}(t)=\Omega_{j} e^{-(t+t_j)^2/2\tau^2}$ for $j=1,2$ with $\Omega_1/2\pi = 30~\rm MHz,$
$\Omega_2/2\pi = 40~\rm MHz,$
$t_1=3.5~\mu\rm s$, $t_2=5.5~\mu\rm s$, $\tau=1~\mu\rm s,$ and $\delta/2\pi= 200~\rm MHz$. The single ARP pulse used $\Omega_{0}(t)=\Omega_{0} e^{-t^2/2\tau^2}$ with $\Omega_0/2\pi=2~\rm MHz$, $\tau=1~\mu\rm s$, and linear chirp $\alpha/2\pi=(1/2\pi)(d\delta(t)/dt)=1 ~{\rm MHz}/\mu\rm s$~\cite{Beterov2011}.
}
\end{figure}

We have studied the population and phase dynamics of the collective states of the atomic ensemble interacting with laser
radiation. Calculations were performed using the Schr\"odinger equation, neglecting spontaneous emission, and assuming
perfect blockade so only states with at most a single Rydberg excitation were included. The details of our calculations
are discussed in Supplemental material. At the end of a double STIRAP sequence the population is returned back to the
collective ground state $\left| {000...} \right\rangle $ of the atomic ensemble, but a geometric phase is accumulated.
This  phase shift of the ground state is dependent on the Rabi frequency and leads to gate errors. We have found that
the phase of the atomic wavefunction can be compensated by switching the sign of the detuning between two STIRAP
pulses, or by switching the phase between two ARP pulses, as shown in Fig.~\ref{Phases}. For a double STIRAP sequence
with the same detuning throughout the accumulated phase depends on $N$ [Fig.~\ref{Phases}(a)], while the phase change
is zero, independent of $N$, when we switch the sign of detuning $\delta$ between the two STIRAP sequences
[Fig.~\ref{Phases}(b)]. A similar phase cancellation  occurs for $\pi$ phase shifted  ARP pulses
[Fig.~\ref{Phases}(c)], which can be implemented using an acousto-optic modulator~\cite{ARPnote}.

The probability of loading $N$ noninteracting atoms in a small  optical or magnetic trap is described, in general, by
Poissonian statistics. For $\bar{N}=5$ the probability to load zero atoms is 0.0067, which is  small enough to create a
large quantum register with a small number of defects~\cite{loadingnote}. Figure~3(a) shows a comparison of the
fidelity of single-atom excitation for a single-photon $\pi$ rotation with the area optimized for $N=5$ atoms compared
to STIRAP or ARP pulses. We see that the adiabatic pulses reduce the population error by up to several orders of
magnitude for a wide range of $N$. Finite lifetimes of the intermediate excited state  and Rydberg states can lead,
however, to breakdown of the deterministic excitation. Figure~3(b) shows the population errors for a single STIRAP
sequence in the ensemble of $N=1-4$ atoms with linewidth of the intermediate state $\gamma_e/(2\pi)=5$~MHz and of the
Rydberg state $\gamma_r/(2\pi)=0.8$~kHz calculated using the  density matrix equations for an ensemble of interacting
atoms~\cite{Petrosyan2013} for two different detunings from the intermediate state $\delta=200$~MHz ($\Omega_1/2\pi =
30~\rm MHz,$ $\Omega_2/2\pi = 40~\rm MHz$, $\tau=1~\mu\rm s$) and $\delta=2$~GHz ($\Omega_1/2\pi = 250~\rm MHz,$
$\Omega_2/2\pi = 250~\rm MHz$, $\tau=0.2~\mu\rm s$,). We see that the effects of the finite lifetime of the
intermediate state are negligible if the detuning from the intermediate state is chosen so that  $\Delta >> \Omega$.

\begin{figure}[!t]
\includegraphics[width=\columnwidth]{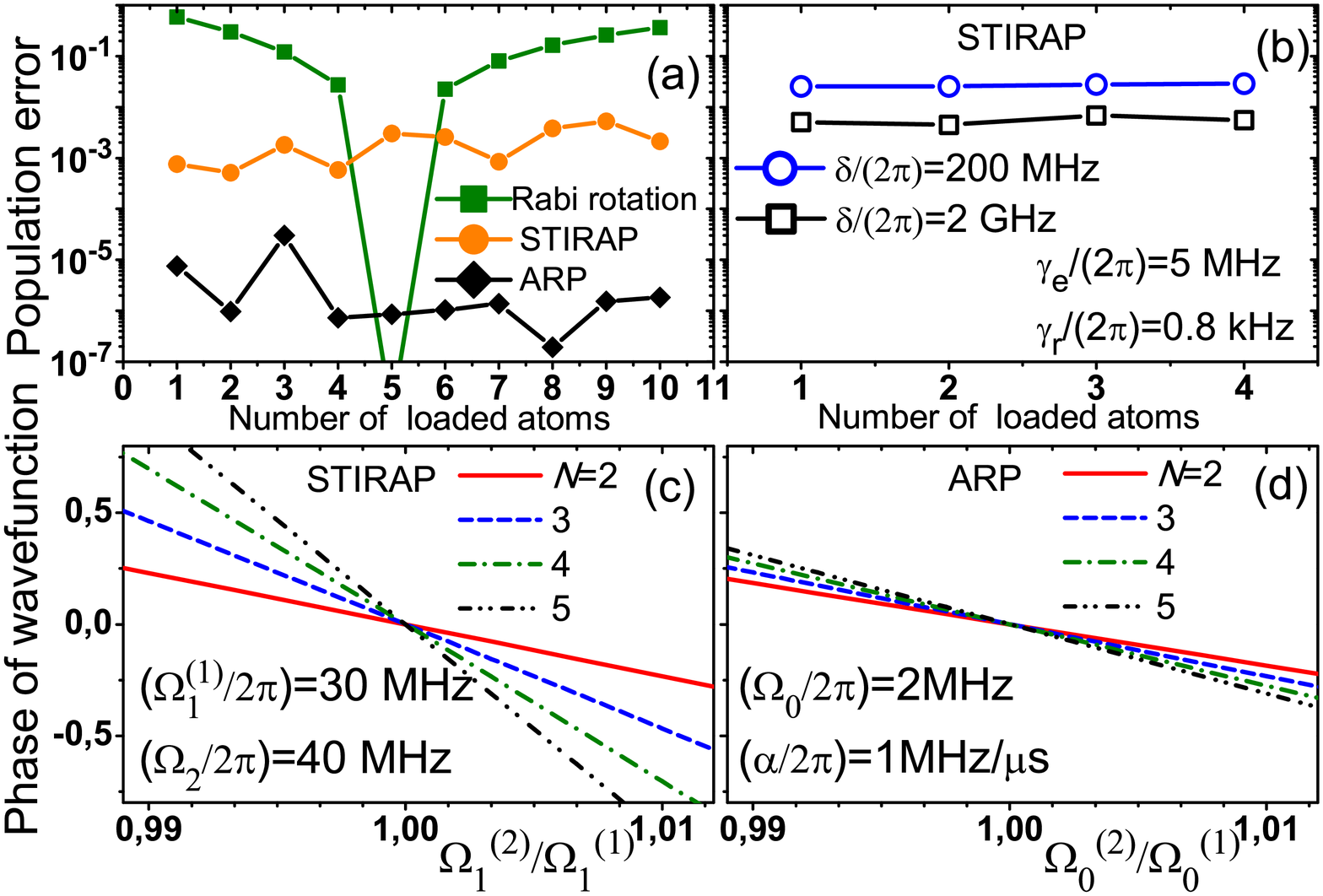}
\vspace{-.8cm}
\caption{
\label{Errors}(Color online).
(a)  Comparison of the fidelity of
single-atom excitation by a $\pi$ laser pulse  having the area optimized for $N=5$ atoms ($t=\pi/\sqrt 5\Omega$), with a  STIRAP sequence, and with an ARP pulse.  All parameters are as in Fig.~\ref{Phases}. Spontaneous emission is not taken into account.
(b) The population error for single STIRAP sequence calculated taking into account  linewidth $\gamma/(2\pi)=5$~MHz of the intermediate state for detuning  $\delta=200$~MHz ($\Omega_1/2\pi = 30~\rm MHz,$ $\Omega_2/2\pi = 40~\rm MHz$, $\tau=1~\mu\rm s$) and $\delta=2$~GHz ($\Omega_1/2\pi = 250~\rm MHz,$ $\Omega_2/2\pi = 250~\rm MHz$, $\tau=0.2~\mu\rm s$); (c),(d) Dependence of the phase error on Rabi frequency changes between pulses for STIRAP or ARP pulses calculated using Schr\"odinger equation.
}
\end{figure}

Although the proposed double-pulse sequences are almost insensitive
 to moderate variations of the absolute Rabi frequency, the main sources of errors are fluctuations of the Rabi frequencies between
 the first and second pulses. For perfectly identical pulses the population transfer error in ensembles of $N=5$ atoms can be kept below $10^{-3}$ for STIRAP and below $10^{-4}$ for
 an ARP pulse  for a wide range of Rabi frequencies.
The dependence  of the phase errors on parameters of the laser pulses are shown in Fig.~\ref{Errors}(c),(d). The
dependence of the phase error on the ratio of Rabi frequencies $\Omega_1^{(2)}/\Omega_1^{(1)}$ between pulses [see
Fig.~\ref{LevelScheme}(b)] is shown in Fig.~\ref{Errors}(c) for $N=1-5$ atoms.  The single-photon ARP excitation  in
Fig.~\ref{Errors}(d) demonstrates reduced sensitivity to  fluctuations of the Rabi frequency and has higher efficiency
at lower Rabi frequencies. Although this could be an important advantage over STIRAP, implementation of single-photon
Rydberg excitation is difficult due to the need of ultraviolet laser radiation and larger sensitivity to Doppler
broadening~\cite{Ryabtsev2011}. For either approach the double pulse amplitudes  must be well matched for low phase
errors. Using a fiber delay line amplitude matching at the level of $10^{-6}$ is feasible over the timescale of few
microseconds~\cite{Wineland1998}.

\begin{figure}
\includegraphics[width=\columnwidth]{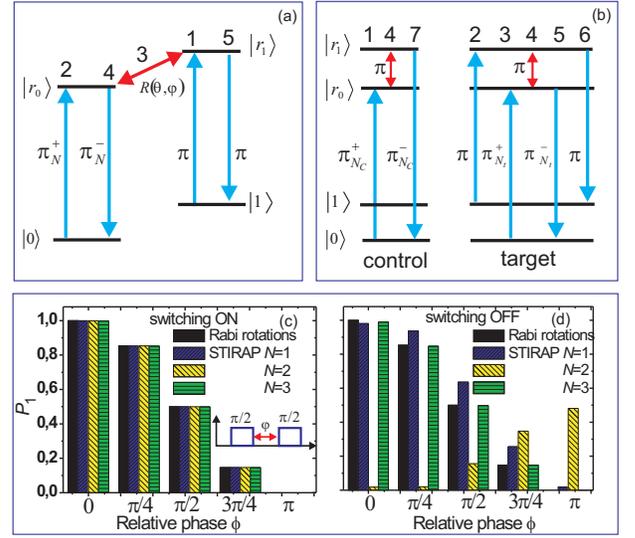}
\vspace{-.6cm}
\caption{
(Color online). (a) Single qubit  gate for a mesoscopic  qubit with $N$ atoms. Pulses $1-5$ act between the qubit states $\ket{0},\ket{1}$ and the Rydberg states $\ket{r_0},\ket{r_1}$.   Pulses $1,2,4,5$ are optical transitions and pulse 3 is a microwave frequency transition between Rydberg states. (b)
CNOT gate between mesoscopic  qubits with $N_c$ atoms in the control qubit and $N_t$ atoms in the target qubit.
(c),(d) The dependence of the population of the qubit state $\ket{1}$ after two sequential $\pi/2$ rotations on the phase difference $\phi$ between the pulses with  (c) and without (d) switching the sign of the detuning between the STIRAP sequences.
\label{fig.scheme}}
\end{figure}

\textit{Gates} We have developed protocols to implement quantum logic gates using phase compensated double STIRAP or
ARP. Consider atoms with levels $\ket{0},\ket{1},\ket{e}\ket{r}$ as shown in Fig.~1. A qubit can be encoded in an $N$
atom ensemble with the logical states $ \ket{\bar 0} = \ket{000...000}$, $\ket{\bar 1}' =
\frac{1}{\sqrt{N}}\sum_{j=1}^N \ket{000 ... 1_j ...000}$. Levels $\ket{0},\ket{1}$ are atomic hyperfine ground states.
Coupling between these states is mediated by the singly excited Rydberg state $\ket{\bar r}' =
\frac{1}{\sqrt{N}}\sum_{j=1}^N \ket{000 ... r_j ...000}.$ Rydberg blockade only allows single excitation of  $\ket{r}$
so the states $\ket{\bar 0}$ and $\ket{\bar r }'$ experience a collectively enhanced coupling rate $\Omega_N=\sqrt
N\Omega$. States $\ket{\bar r}'$ and $\ket{\bar 1 }'$ are  coupled at the single atom rate $\Omega$. State $\ket{\bar
1}'$ is produced by the sequential application of $\pi$ pulses $\ket{\bar 0}\rightarrow\ket{\bar r}'$ and $\ket{\bar
r}'\rightarrow\ket{\bar 1}'$.

  Pulse areas independent of $N$ on the $\ket{0} \leftrightarrow \ket{r}$ transition can be implemented with STIRAP or ARP
 as described above.
 We will define the logical basis states and the auxiliary Rydberg state as $\ket{\bar 0} = \ket{000...000},$
$\ket{\bar 1} = e^{\imath\chi_N} \ket{\bar 1}',$ and $\ket{\bar r} = e^{\imath\chi_N}\ket{\bar r}'.$ Here $\chi_N$ is
the phase produced by a single $N$ atom STIRAP pulse with positive detuning.  We assume that we do not know the value
of $N$, which may vary from qubit to qubit,  and therefore $\chi_N$ is also unknown, but has a definite value for fixed
$N$.

We find that arbitrary single qubit rotations in the basis $\ket{\bar 0},\ket{\bar 1}$ can be performed with high
fidelity, without precise knowledge of $N$, by accessing several Rydberg levels $\ket{r_0}$, $\ket{r_1}$ as shown in
Fig.~\ref{fig.scheme}(a). The equations which describe the gate sequence are discussed in Supplemental material. The
final state $\ket{\psi}=a'\ket{\bar 0} + b' \ket{\bar 1}$ is arbitrary and is selected by the rotation
$R(\theta,\phi)$,  in step 3:
 $\begin{pmatrix}a'\\-b'\end{pmatrix}={\bf R}(\theta,\phi)\begin{pmatrix}a\\b\end{pmatrix}$. Depending on the choice of
  implementation, to be discussed below,  this may be given by a one- or two-photon microwave pulse, with Rabi frequency $\Omega_3$.
  Provided  states $\ket{r_0}, \ket{r_1}$ are strongly interacting, and
  limit the number of excitations in the  ensemble to one, the indicated sequence is obtained.  In the regime of $\Omega_3$ large compared
   to the Rydberg excitation rates the time spent populating a Rydberg level corresponds to $4\pi$ of Rydberg pulse area.
   This is the same as for a single atom
$C_Z$ gate, and  we  therefore expect the limit on gate infidelity to be $\sim 0.002$~\cite{XZhang2012} for small
ensembles. It was shown in ~\cite{Saffman2008} that in a 3D lattice the number of atoms $N$ which can be entangled at
fixed error scales linearly. Although the details of the error scaling are different for ensemble qubits, for moderate
size ensembles we anticipate approximately linear scaling, with a numerical prefactor that requires a detailed analysis
to be given elsewhere.

The five pulse sequence we describe here is more complicated than the three pulses needed for an arbitrary  single
qubit gate in the approach of Ref.~\cite{Lukin2001}. The reason for this added complexity is that the special phase
preserving property of the double STIRAP or ARP sequences requires that all population is initially in one of the
states connected by the  pulses. The sequence of pulses in Fig.~\ref{fig.scheme}(a) ensures that this condition is
always satisfied.

To verify that our scheme preserves coherence, we have numerically modelled the sequence of two single-qubit rotations
for an angle of $\pi/2$ with relative phases $\phi$ in the range $0-\pi$. The probability to find the ensemble in the
qubit state $\ket{1}$  was calculated for our STIRAP-based protocol for $N=1-3$ atoms  and compared with the outcome of
similar single-atom gate sequence applied using conventional Rabi rotations [shown as black in Fig.~4(c)]. We have
found that the probability for the ensemble to be in state $\ket{1}$ is independent of the number of atoms and
correctly depends on the relative phase between the microwave pulses, as shown in Fig.~4(c). In contrast, if we do not
switch the detuning from the intermediate state after the first STIRAP pulse, the probability to find the ensemble in
the state $\ket{1}$  becomes $N$-dependent and is inconsistent with the expected values, as shown in Fig.~4(d).

A CNOT gate  can be implemented by the sequence H(t) - $C_Z$ - H(t) ~\cite{Nielsen2000}, where the Hadamard gates are
performed as in Fig.~\ref{fig.scheme}(a). The $C_Z$ operation is implemented in analogy to schemes for single atom
qubits~\cite{Jaksch2000} mediated by Rydberg interactions, using the protocol $\pi_{\ket{\bar 1}-\ket{\bar r}}({\rm
c})$ $2\pi_{\ket{\bar 1}-\ket{\bar r}}({\rm t})$ $\pi_{\ket{\bar 1}-\ket{\bar r}}({\rm c})$, where c(t) stand for
control(target) qubits. The CNOT gate therefore requires a total pulse area of $12\pi$ Rydberg pulses. We can reduce
this to $7\pi$ of Rydberg pulses as shown in Fig.~\ref{fig.scheme}(b) which implements an approach analogous to the
amplitude-swap gate demonstrated for single atom qubits in~\cite{Isenhower2010}. All pulses except number 4 in the
sequence are optical and are localized to either the control or target qubit. Pulse 4 is a microwave field and drives a
$\pi$ rotation on both qubits. As for the single qubit gate the requirement for high fidelity operation is that the
interactions $\ket{r_0} \leftrightarrow \ket{r_0}$, $\ket{r_1} \leftrightarrow \ket{r_1}$, $\ket{r_0} \leftrightarrow
\ket{r_1}$ all lead to full blockade of the ensembles, and we refer to the supplemental material for the choice of $n$
that fullfill this condition. Since the frequency of pulse 4, which is determined by the energy separation of states
$\ket{r_0},\ket{r_1}$, can be chosen to be very different from the qubit frequency given by the energy separation of
states $\ket{0},\ket{1}$ the application of microwave pulses will not lead to crosstalk in an array of ensemble qubits.

In summary we have demonstrated that  double STIRAP and ARP sequences with phase compensation enable high fidelity
quantum gates in collectively encoded ensembles.  We have shown that phase compensation using this method works
effectively regardless of the number of atoms $N$ even in small atomic ensembles randomly loaded, which display a large
fractional variation in $N$. We have presented full protocols for one-qubit and two-qubit logic gates which perform at
high fidelity both in the regime of small and large ensembles. We anticipate that these ideas will contribute to
realization of quantum logic using collectively encoded qubits and registers.

\begin{acknowledgments}
This work was supported by the grant of the President of Russian Federation MK.7060.2012.2, EPSRC project EP/K022938/1, RAS, RFBR and Russian Quantum Center.  MS was supported by the NSF and  the AFOSR MURI program.
\end{acknowledgments}


\begin{thebibliography}{33}%
\makeatletter
\providecommand \@ifxundefined [1]{%
 \@ifx{#1\undefined}
}%
\providecommand \@ifnum [1]{%
 \ifnum #1\expandafter \@firstoftwo
 \else \expandafter \@secondoftwo
 \fi
}%
\providecommand \@ifx [1]{%
 \ifx #1\expandafter \@firstoftwo
 \else \expandafter \@secondoftwo
 \fi
}%
\providecommand \natexlab [1]{#1}%
\providecommand \enquote  [1]{``#1''}%
\providecommand \bibnamefont  [1]{#1}%
\providecommand \bibfnamefont [1]{#1}%
\providecommand \citenamefont [1]{#1}%
\providecommand \href@noop [0]{\@secondoftwo}%
\providecommand \href [0]{\begingroup \@sanitize@url \@href}%
\providecommand \@href[1]{\@@startlink{#1}\@@href}%
\providecommand \@@href[1]{\endgroup#1\@@endlink}%
\providecommand \@sanitize@url [0]{\catcode `\\12\catcode `\$12\catcode
  `\&12\catcode `\#12\catcode `\^12\catcode `\_12\catcode `\%12\relax}%
\providecommand \@@startlink[1]{}%
\providecommand \@@endlink[0]{}%
\providecommand \url  [0]{\begingroup\@sanitize@url \@url }%
\providecommand \@url [1]{\endgroup\@href {#1}{\urlprefix }}%
\providecommand \urlprefix  [0]{URL }%
\providecommand \Eprint [0]{\href }%
\providecommand \doibase [0]{http://dx.doi.org/}%
\providecommand \selectlanguage [0]{\@gobble}%
\providecommand \bibinfo  [0]{\@secondoftwo}%
\providecommand \bibfield  [0]{\@secondoftwo}%
\providecommand \translation [1]{[#1]}%
\providecommand \BibitemOpen [0]{}%
\providecommand \bibitemStop [0]{}%
\providecommand \bibitemNoStop [0]{.\EOS\space}%
\providecommand \EOS [0]{\spacefactor3000\relax}%
\providecommand \BibitemShut  [1]{\csname bibitem#1\endcsname}%
\let\auto@bib@innerbib\@empty
\bibitem [{\citenamefont {Lukin}\ \emph {et~al.}(2001)\citenamefont {Lukin},
  \citenamefont {Fleischhauer}, \citenamefont {Cote}, \citenamefont {Duan},
  \citenamefont {Jaksch}, \citenamefont {Cirac},\ and\ \citenamefont
  {Zoller}}]{Lukin2001}%
  \BibitemOpen
  \bibfield  {author} {\bibinfo {author} {\bibfnamefont {M.~D.}\ \bibnamefont
  {Lukin}}, \bibinfo {author} {\bibfnamefont {M.}~\bibnamefont {Fleischhauer}},
  \bibinfo {author} {\bibfnamefont {R.}~\bibnamefont {Cote}}, \bibinfo {author}
  {\bibfnamefont {L.~M.}\ \bibnamefont {Duan}}, \bibinfo {author}
  {\bibfnamefont {D.}~\bibnamefont {Jaksch}}, \bibinfo {author} {\bibfnamefont
  {J.~I.}\ \bibnamefont {Cirac}}, \ and\ \bibinfo {author} {\bibfnamefont
  {P.}~\bibnamefont {Zoller}},\ }\href@noop {} {\bibfield  {journal} {\bibinfo
  {journal} {Phys. Rev. Lett.}\ }\textbf {\bibinfo {volume} {87}},\ \bibinfo
  {pages} {037901} (\bibinfo {year} {2001})}\BibitemShut {NoStop}%
\bibitem [{\citenamefont {Brion}\ \emph {et~al.}(2007)\citenamefont {Brion},
  \citenamefont {M\o{}lmer},\ and\ \citenamefont {Saffman}}]{Brion2007d}%
  \BibitemOpen
  \bibfield  {author} {\bibinfo {author} {\bibfnamefont {E.}~\bibnamefont
  {Brion}}, \bibinfo {author} {\bibfnamefont {K.}~\bibnamefont {M\o{}lmer}}, \
  and\ \bibinfo {author} {\bibfnamefont {M.}~\bibnamefont {Saffman}},\
  }\href@noop {} {\bibfield  {journal} {\bibinfo  {journal} {Phys. Rev. Lett.}\
  }\textbf {\bibinfo {volume} {99}},\ \bibinfo {pages} {260501} (\bibinfo
  {year} {2007})}\BibitemShut {NoStop}%
\bibitem [{\citenamefont {Saffman}\ and\ \citenamefont
  {M\o{}lmer}(2008)}]{Saffman2008}%
  \BibitemOpen
  \bibfield  {author} {\bibinfo {author} {\bibfnamefont {M.}~\bibnamefont
  {Saffman}}\ and\ \bibinfo {author} {\bibfnamefont {K.}~\bibnamefont
  {M\o{}lmer}},\ }\href@noop {} {\bibfield  {journal} {\bibinfo  {journal}
  {Phys. Rev. A}\ }\textbf {\bibinfo {volume} {78}},\ \bibinfo {pages} {012336}
  (\bibinfo {year} {2008})}\BibitemShut {NoStop}%
\bibitem [{\citenamefont {Whitlock}\ \emph {et~al.}(2009)\citenamefont
  {Whitlock}, \citenamefont {Gerritsma}, \citenamefont {Fernholz},\ and\
  \citenamefont {Spreeuw}}]{Whitlock2009}%
  \BibitemOpen
  \bibfield  {author} {\bibinfo {author} {\bibfnamefont {S.}~\bibnamefont
  {Whitlock}}, \bibinfo {author} {\bibfnamefont {R.}~\bibnamefont {Gerritsma}},
  \bibinfo {author} {\bibfnamefont {T.}~\bibnamefont {Fernholz}}, \ and\
  \bibinfo {author} {\bibfnamefont {R.~J.~C.}\ \bibnamefont {Spreeuw}},\
  }\href@noop {} {\bibfield  {journal} {\bibinfo  {journal} {New J. Phys.}\
  }\textbf {\bibinfo {volume} {11}},\ \bibinfo {pages} {023021} (\bibinfo
  {year} {2009})}\BibitemShut {NoStop}%
\bibitem [{\citenamefont {Tordrup}\ \emph {et~al.}(2008)\citenamefont
  {Tordrup}, \citenamefont {Negretti},\ and\ \citenamefont
  {M\o{}lmer}}]{Tordrup2008}%
  \BibitemOpen
  \bibfield  {author} {\bibinfo {author} {\bibfnamefont {K.}~\bibnamefont
  {Tordrup}}, \bibinfo {author} {\bibfnamefont {A.}~\bibnamefont {Negretti}}, \
  and\ \bibinfo {author} {\bibfnamefont {K.}~\bibnamefont {M\o{}lmer}},\
  }\href@noop {} {\bibfield  {journal} {\bibinfo  {journal} {Phys. Rev. Lett.}\
  }\textbf {\bibinfo {volume} {101}},\ \bibinfo {pages} {040501} (\bibinfo
  {year} {2008})}\BibitemShut {NoStop}%
\bibitem [{\citenamefont {Wesenberg}\ \emph {et~al.}(2009)\citenamefont
  {Wesenberg}, \citenamefont {Ardavan}, \citenamefont {Briggs}, \citenamefont
  {Morton}, \citenamefont {Schoelkopf}, \citenamefont {Schuster},\ and\
  \citenamefont {M\o{}lmer}}]{Wesenberg2009}%
  \BibitemOpen
  \bibfield  {author} {\bibinfo {author} {\bibfnamefont {J.~H.}\ \bibnamefont
  {Wesenberg}}, \bibinfo {author} {\bibfnamefont {A.}~\bibnamefont {Ardavan}},
  \bibinfo {author} {\bibfnamefont {G.~A.~D.}\ \bibnamefont {Briggs}}, \bibinfo
  {author} {\bibfnamefont {J.~J.~L.}\ \bibnamefont {Morton}}, \bibinfo {author}
  {\bibfnamefont {R.~J.}\ \bibnamefont {Schoelkopf}}, \bibinfo {author}
  {\bibfnamefont {D.~I.}\ \bibnamefont {Schuster}}, \ and\ \bibinfo {author}
  {\bibfnamefont {K.}~\bibnamefont {M\o{}lmer}},\ }\href@noop {} {\bibfield
  {journal} {\bibinfo  {journal} {Phys. Rev. Lett.}\ }\textbf {\bibinfo
  {volume} {103}},\ \bibinfo {pages} {070502} (\bibinfo {year}
  {2009})}\BibitemShut {NoStop}%
\bibitem [{\citenamefont {Dudin}\ and\ \citenamefont
  {Kuzmich}(2012)}]{Dudin2012}%
  \BibitemOpen
  \bibfield  {author} {\bibinfo {author} {\bibfnamefont {Y.~O.}\ \bibnamefont
  {Dudin}}\ and\ \bibinfo {author} {\bibfnamefont {A.}~\bibnamefont
  {Kuzmich}},\ }\href@noop {} {\bibfield  {journal} {\bibinfo  {journal}
  {Science}\ }\textbf {\bibinfo {volume} {336}},\ \bibinfo {pages} {887}
  (\bibinfo {year} {2012})}\BibitemShut {NoStop}%
\bibitem [{\citenamefont {Peyronel}\ \emph {et~al.}(2012)\citenamefont
  {Peyronel}, \citenamefont {Firstenberg}, \citenamefont {Liang}, \citenamefont
  {Hofferberth}, \citenamefont {Gorshkov}, \citenamefont {Pohl}, \citenamefont
  {Lukin},\ and\ \citenamefont {Vuleti\'c}}]{Peyronel2012}%
  \BibitemOpen
  \bibfield  {author} {\bibinfo {author} {\bibfnamefont {T.}~\bibnamefont
  {Peyronel}}, \bibinfo {author} {\bibfnamefont {O.}~\bibnamefont
  {Firstenberg}}, \bibinfo {author} {\bibfnamefont {Q.-Y.}\ \bibnamefont
  {Liang}}, \bibinfo {author} {\bibfnamefont {S.}~\bibnamefont {Hofferberth}},
  \bibinfo {author} {\bibfnamefont {A.~V.}\ \bibnamefont {Gorshkov}}, \bibinfo
  {author} {\bibfnamefont {T.}~\bibnamefont {Pohl}}, \bibinfo {author}
  {\bibfnamefont {M.~D.}\ \bibnamefont {Lukin}}, \ and\ \bibinfo {author}
  {\bibfnamefont {V.}~\bibnamefont {Vuleti\'c}},\ }\href@noop {} {\bibfield
  {journal} {\bibinfo  {journal} {Nature}\ }\textbf {\bibinfo {volume} {488}},\
  \bibinfo {pages} {57} (\bibinfo {year} {2012})}\BibitemShut {NoStop}%
\bibitem [{\citenamefont {Jiang}\ \emph {et~al.}(2007)\citenamefont {Jiang},
  \citenamefont {Taylor}, \citenamefont {S\o{}rensen},\ and\ \citenamefont
  {Lukin}}]{Jiang2007}%
  \BibitemOpen
  \bibfield  {author} {\bibinfo {author} {\bibfnamefont {L.}~\bibnamefont
  {Jiang}}, \bibinfo {author} {\bibfnamefont {J.~M.}\ \bibnamefont {Taylor}},
  \bibinfo {author} {\bibfnamefont {A.~S.}\ \bibnamefont {S\o{}rensen}}, \ and\
  \bibinfo {author} {\bibfnamefont {M.~D.}\ \bibnamefont {Lukin}},\ }\href@noop
  {} {\bibfield  {journal} {\bibinfo  {journal} {Phys. Rev. A}\ }\textbf
  {\bibinfo {volume} {76}},\ \bibinfo {pages} {062323} (\bibinfo {year}
  {2007})}\BibitemShut {NoStop}%
\bibitem [{\citenamefont {Pedersen}\ and\ \citenamefont
  {M\o{}lmer}(2009)}]{Pedersen2009}%
  \BibitemOpen
  \bibfield  {author} {\bibinfo {author} {\bibfnamefont {L.~H.}\ \bibnamefont
  {Pedersen}}\ and\ \bibinfo {author} {\bibfnamefont {K.}~\bibnamefont
  {M\o{}lmer}},\ }\href@noop {} {\bibfield  {journal} {\bibinfo  {journal}
  {Phys. Rev. A}\ }\textbf {\bibinfo {volume} {79}},\ \bibinfo {pages} {012320}
  (\bibinfo {year} {2009})}\BibitemShut {NoStop}%
\bibitem [{\citenamefont {Saffman}\ and\ \citenamefont
  {Walker}(2002)}]{Saffman2002}%
  \BibitemOpen
  \bibfield  {author} {\bibinfo {author} {\bibfnamefont {M.}~\bibnamefont
  {Saffman}}\ and\ \bibinfo {author} {\bibfnamefont {T.~G.}\ \bibnamefont
  {Walker}},\ }\href@noop {} {\bibfield  {journal} {\bibinfo  {journal} {Phys.
  Rev. A}\ }\textbf {\bibinfo {volume} {66}},\ \bibinfo {pages} {065403}
  (\bibinfo {year} {2002})}\BibitemShut {NoStop}%
\bibitem [{\citenamefont {M\"uller}\ \emph {et~al.}(2009)\citenamefont
  {M\"uller}, \citenamefont {Lesanovsky}, \citenamefont {Weimer}, \citenamefont
  {B\"uchler},\ and\ \citenamefont {Zoller}}]{Muller2009}%
  \BibitemOpen
  \bibfield  {author} {\bibinfo {author} {\bibfnamefont {M.}~\bibnamefont
  {M\"uller}}, \bibinfo {author} {\bibfnamefont {I.}~\bibnamefont
  {Lesanovsky}}, \bibinfo {author} {\bibfnamefont {H.}~\bibnamefont {Weimer}},
  \bibinfo {author} {\bibfnamefont {H.~P.}\ \bibnamefont {B\"uchler}}, \ and\
  \bibinfo {author} {\bibfnamefont {P.}~\bibnamefont {Zoller}},\ }\href@noop {}
  {\bibfield  {journal} {\bibinfo  {journal} {Phys. Rev. Lett.}\ }\textbf
  {\bibinfo {volume} {102}},\ \bibinfo {pages} {170502} (\bibinfo {year}
  {2009})}\BibitemShut {NoStop}%
\bibitem [{\citenamefont {Saffman}\ and\ \citenamefont
  {M\o{}lmer}(2009)}]{Saffman2009b}%
  \BibitemOpen
  \bibfield  {author} {\bibinfo {author} {\bibfnamefont {M.}~\bibnamefont
  {Saffman}}\ and\ \bibinfo {author} {\bibfnamefont {K.}~\bibnamefont
  {M\o{}lmer}},\ }\href@noop {} {\bibfield  {journal} {\bibinfo  {journal}
  {Phys. Rev. Lett.}\ }\textbf {\bibinfo {volume} {102}},\ \bibinfo {pages}
  {240502} (\bibinfo {year} {2009})}\BibitemShut {NoStop}%
\bibitem [{\citenamefont {Isenhower}\ \emph {et~al.}(2011)\citenamefont
  {Isenhower}, \citenamefont {Saffman},\ and\ \citenamefont
  {M\o{}lmer}}]{Isenhower2011}%
  \BibitemOpen
  \bibfield  {author} {\bibinfo {author} {\bibfnamefont {L.}~\bibnamefont
  {Isenhower}}, \bibinfo {author} {\bibfnamefont {M.}~\bibnamefont {Saffman}},
  \ and\ \bibinfo {author} {\bibfnamefont {K.}~\bibnamefont {M\o{}lmer}},\
  }\href@noop {} {\bibfield  {journal} {\bibinfo  {journal} {Quant. Inf.
  Proc.}\ }\textbf {\bibinfo {volume} {10}},\ \bibinfo {pages} {755} (\bibinfo
  {year} {2011})}\BibitemShut {NoStop}%
\bibitem [{\citenamefont {Saffman}\ \emph {et~al.}(2010)\citenamefont
  {Saffman}, \citenamefont {Walker},\ and\ \citenamefont
  {M\o{}lmer}}]{Saffman2010}%
  \BibitemOpen
  \bibfield  {author} {\bibinfo {author} {\bibfnamefont {M.}~\bibnamefont
  {Saffman}}, \bibinfo {author} {\bibfnamefont {T.~G.}\ \bibnamefont {Walker}},
  \ and\ \bibinfo {author} {\bibfnamefont {K.}~\bibnamefont {M\o{}lmer}},\
  }\href@noop {} {\bibfield  {journal} {\bibinfo  {journal} {Rev. Mod. Phys.}\
  }\textbf {\bibinfo {volume} {82}},\ \bibinfo {pages} {2313} (\bibinfo {year}
  {2010})}\BibitemShut {NoStop}%
\bibitem [{\citenamefont {Whitlock}\ \emph {et~al.}(2010)\citenamefont
  {Whitlock}, \citenamefont {Ockeloen},\ and\ \citenamefont
  {Spreeuw}}]{Whitlock2010b}%
  \BibitemOpen
  \bibfield  {author} {\bibinfo {author} {\bibfnamefont {S.}~\bibnamefont
  {Whitlock}}, \bibinfo {author} {\bibfnamefont {C.~F.}\ \bibnamefont
  {Ockeloen}}, \ and\ \bibinfo {author} {\bibfnamefont {R.~J.~C.}\ \bibnamefont
  {Spreeuw}},\ }\href@noop {} {\bibfield  {journal} {\bibinfo  {journal} {Phys.
  Rev. Lett.}\ }\textbf {\bibinfo {volume} {104}},\ \bibinfo {pages} {120402}
  (\bibinfo {year} {2010})}\BibitemShut {NoStop}%
\bibitem [{\citenamefont {Zhang}\ \emph
  {et~al.}(2012{\natexlab{a}})\citenamefont {Zhang}, \citenamefont {McConnell},
  \citenamefont {\ifmmode~\acute{C}\else \'{C}\fi{}uk}, \citenamefont {Lin},
  \citenamefont {Schleier-Smith}, \citenamefont {Leroux},\ and\ \citenamefont
  {Vuleti\ifmmode~\acute{c}\else \'{c}\fi{}}}]{HZhang2012}%
  \BibitemOpen
  \bibfield  {author} {\bibinfo {author} {\bibfnamefont {H.}~\bibnamefont
  {Zhang}}, \bibinfo {author} {\bibfnamefont {R.}~\bibnamefont {McConnell}},
  \bibinfo {author} {\bibfnamefont {S.}~\bibnamefont {\ifmmode~\acute{C}\else
  \'{C}\fi{}uk}}, \bibinfo {author} {\bibfnamefont {Q.}~\bibnamefont {Lin}},
  \bibinfo {author} {\bibfnamefont {M.~H.}\ \bibnamefont {Schleier-Smith}},
  \bibinfo {author} {\bibfnamefont {I.~D.}\ \bibnamefont {Leroux}}, \ and\
  \bibinfo {author} {\bibfnamefont {V.}~\bibnamefont
  {Vuleti\ifmmode~\acute{c}\else \'{c}\fi{}}},\ }\href@noop {} {\bibfield
  {journal} {\bibinfo  {journal} {Phys. Rev. Lett.}\ }\textbf {\bibinfo
  {volume} {109}},\ \bibinfo {pages} {133603} (\bibinfo {year}
  {2012}{\natexlab{a}})}\BibitemShut {NoStop}%
\bibitem [{\citenamefont {Bergmann}\ \emph {et~al.}(1998)\citenamefont
  {Bergmann}, \citenamefont {Theuer},\ and\ \citenamefont
  {Shore}}]{Bergmann1998}%
  \BibitemOpen
  \bibfield  {author} {\bibinfo {author} {\bibfnamefont {K.}~\bibnamefont
  {Bergmann}}, \bibinfo {author} {\bibfnamefont {H.}~\bibnamefont {Theuer}}, \
  and\ \bibinfo {author} {\bibfnamefont {B.~W.}\ \bibnamefont {Shore}},\
  }\href@noop {} {\bibfield  {journal} {\bibinfo  {journal} {Rev. Mod. Phys.}\
  }\textbf {\bibinfo {volume} {70}},\ \bibinfo {pages} {1003} (\bibinfo {year}
  {1998})}\BibitemShut {NoStop}%
\bibitem [{\citenamefont {Abragam}(1961)}]{Abragam1961}%
  \BibitemOpen
  \bibfield  {author} {\bibinfo {author} {\bibfnamefont {A.}~\bibnamefont
  {Abragam}},\ }\href@noop {} {\emph {\bibinfo {title} {Principles of Nuclear
  Magnetism}}}\ (\bibinfo  {publisher} {Oxford University Press},\ \bibinfo
  {year} {1961})\BibitemShut {NoStop}%
\bibitem [{\citenamefont {Marte}\ \emph {et~al.}(1991)\citenamefont {Marte},
  \citenamefont {Zoller},\ and\ \citenamefont {Hall}}]{Marte1991}%
  \BibitemOpen
  \bibfield  {author} {\bibinfo {author} {\bibfnamefont {P.}~\bibnamefont
  {Marte}}, \bibinfo {author} {\bibfnamefont {P.}~\bibnamefont {Zoller}}, \
  and\ \bibinfo {author} {\bibfnamefont {J.~L.}\ \bibnamefont {Hall}},\
  }\href@noop {} {\bibfield  {journal} {\bibinfo  {journal} {Phys. Rev. A}\
  }\textbf {\bibinfo {volume} {44}},\ \bibinfo {pages} {R4118} (\bibinfo {year}
  {1991})}\BibitemShut {NoStop}%
\bibitem [{\citenamefont {Kis}\ and\ \citenamefont {Renzoni}(2002)}]{Kis2002}%
  \BibitemOpen
  \bibfield  {author} {\bibinfo {author} {\bibfnamefont {Z.}~\bibnamefont
  {Kis}}\ and\ \bibinfo {author} {\bibfnamefont {F.}~\bibnamefont {Renzoni}},\
  }\href@noop {} {\bibfield  {journal} {\bibinfo  {journal} {Phys. Rev. A}\
  }\textbf {\bibinfo {volume} {65}},\ \bibinfo {pages} {032318} (\bibinfo
  {year} {2002})}\BibitemShut {NoStop}%
\bibitem [{\citenamefont {M\o{}ller}\ \emph {et~al.}(2008)\citenamefont
  {M\o{}ller}, \citenamefont {Madsen},\ and\ \citenamefont
  {M\o{}lmer}}]{Moller2008}%
  \BibitemOpen
  \bibfield  {author} {\bibinfo {author} {\bibfnamefont {D.}~\bibnamefont
  {M\o{}ller}}, \bibinfo {author} {\bibfnamefont {L.~B.}\ \bibnamefont
  {Madsen}}, \ and\ \bibinfo {author} {\bibfnamefont {K.}~\bibnamefont
  {M\o{}lmer}},\ }\href@noop {} {\bibfield  {journal} {\bibinfo  {journal}
  {Phys. Rev. Lett.}\ }\textbf {\bibinfo {volume} {100}},\ \bibinfo {pages}
  {170504} (\bibinfo {year} {2008})}\BibitemShut {NoStop}%
\bibitem [{\citenamefont {Pohl}\ \emph {et~al.}(2010)\citenamefont {Pohl},
  \citenamefont {Demler},\ and\ \citenamefont {Lukin}}]{Pohl2010}%
  \BibitemOpen
  \bibfield  {author} {\bibinfo {author} {\bibfnamefont {T.}~\bibnamefont
  {Pohl}}, \bibinfo {author} {\bibfnamefont {E.}~\bibnamefont {Demler}}, \ and\
  \bibinfo {author} {\bibfnamefont {M.~D.}\ \bibnamefont {Lukin}},\ }\href@noop
  {} {\bibfield  {journal} {\bibinfo  {journal} {Phys. Rev. Lett,}\ }\textbf
  {\bibinfo {volume} {104}},\ \bibinfo {pages} {043002} (\bibinfo {year}
  {2010})}\BibitemShut {NoStop}%
\bibitem [{\citenamefont {Beterov}\ \emph {et~al.}(2011)\citenamefont
  {Beterov}, \citenamefont {Tretyakov}, \citenamefont {Entin}, \citenamefont
  {Yakshina}, \citenamefont {Ryabtsev}, \citenamefont {MacCormick},\ and\
  \citenamefont {Bergamini}}]{Beterov2011}%
  \BibitemOpen
  \bibfield  {author} {\bibinfo {author} {\bibfnamefont {I.~I.}\ \bibnamefont
  {Beterov}}, \bibinfo {author} {\bibfnamefont {D.~B.}\ \bibnamefont
  {Tretyakov}}, \bibinfo {author} {\bibfnamefont {V.~M.}\ \bibnamefont
  {Entin}}, \bibinfo {author} {\bibfnamefont {E.~A.}\ \bibnamefont {Yakshina}},
  \bibinfo {author} {\bibfnamefont {I.~I.}\ \bibnamefont {Ryabtsev}}, \bibinfo
  {author} {\bibfnamefont {C.}~\bibnamefont {MacCormick}}, \ and\ \bibinfo
  {author} {\bibfnamefont {S.}~\bibnamefont {Bergamini}},\ }\href@noop {}
  {\bibfield  {journal} {\bibinfo  {journal} {Phys. Rev. A}\ }\textbf {\bibinfo
  {volume} {84}},\ \bibinfo {pages} {023413} (\bibinfo {year}
  {2011})}\BibitemShut {NoStop}%
\bibitem [{loa()}]{loadingnote}%
  \BibitemOpen
  \href@noop {} {}\bibinfo {note} {If the probability of loading at least 1
  atom in a single ensemble is $P_1$, then the probability of successfully
  loading $M$ ensembles after $m$ tries with no defects is $P=1-(1-P_1^M)^m$.
  For $\bar N = 5$ we have $P_1 = 1-0.0067$ and for $M=100$ ensembles the
  probability of a success after $m=10$ tries is 0.999.}\BibitemShut {Stop}%
\bibitem [{ARP()}]{ARPnote}%
  \BibitemOpen
  \href@noop {} {}\bibinfo {note} {If the ARP pulses have the same phase the
  accumulated atomic phase is also independent of $N$, but is equal to
  $\pi$.}\BibitemShut {Stop}%
\bibitem [{\citenamefont {Petrosyan}\ and\ \citenamefont {Molmer}(2002)}]{Petrosyan2013}%
  \BibitemOpen
  \bibfield  {author} {\bibinfo {author} {\bibfnamefont {D.}~\bibnamefont
  {Petrosyan}}, \ and\
  \bibinfo {author} {\bibfnamefont {K.}~\bibnamefont {M\o{}lmer}},\ }\href@noop
  {} {\bibfield  {journal} {\bibinfo  {journal} {arXiv:1302.0682}\ }\textbf
  {\bibinfo {volume} {}},\ \bibinfo {pages} {} (\bibinfo {year}
  {2013})}\BibitemShut {NoStop}%
\bibitem [{\citenamefont {Ryabtsev}\ \emph {et~al.}(2011)\citenamefont
  {Ryabtsev}, \citenamefont {Beterov},\citenamefont {Tretyakov}, \citenamefont {Entin},\ and\ \citenamefont
  {Yakshina}}]{Ryabtsev2011}%
  \BibitemOpen
  \bibfield  {author} {\bibinfo {author} {\bibfnamefont {I.~I.}\ \bibnamefont
  {Ryabtsev}}, \bibinfo {author} {\bibfnamefont {I.~I.}\ \bibnamefont
  {Beterov}}, \bibinfo {author} {\bibfnamefont {D.~B.}\ \bibnamefont
  {Tretyakov}}, \bibinfo {author} {\bibfnamefont {V.~M.}\ \bibnamefont {Entin}},
  \bibinfo {author} {\bibfnamefont {E.~A.}\ \bibnamefont {Yakshina}},\ }\href@noop {}
  {\bibfield  {journal} {\bibinfo  {journal} {Phys. Rev. A}\ }\textbf {\bibinfo
  {volume} {84}},\ \bibinfo {pages} {053409} (\bibinfo {year}
  {2011})}\BibitemShut {NoStop}%
\bibitem [{\citenamefont {Wineland}\ \emph {et~al.}(1998)\citenamefont
  {Wineland}, \citenamefont {Monroe}, \citenamefont {Itano}, \citenamefont
  {Leibfried}, \citenamefont {King},\ and\ \citenamefont
  {Meekhof}}]{Wineland1998}%
  \BibitemOpen
  \bibfield  {author} {\bibinfo {author} {\bibfnamefont {D.~J.}\ \bibnamefont
  {Wineland}}, \bibinfo {author} {\bibfnamefont {C.}~\bibnamefont {Monroe}},
  \bibinfo {author} {\bibfnamefont {W.~M.}\ \bibnamefont {Itano}}, \bibinfo
  {author} {\bibfnamefont {D.}~\bibnamefont {Leibfried}}, \bibinfo {author}
  {\bibfnamefont {B.~E.}\ \bibnamefont {King}}, \ and\ \bibinfo {author}
  {\bibfnamefont {D.~M.}\ \bibnamefont {Meekhof}},\ }\href@noop {} {\bibfield
  {journal} {\bibinfo  {journal} {J. Res. Natl. Inst. Stand. Technol.}\
  }\textbf {\bibinfo {volume} {103}},\ \bibinfo {pages} {259} (\bibinfo {year}
  {1998})}\BibitemShut {NoStop}%
\bibitem [{\citenamefont {Zhang}\ \emph
  {et~al.}(2012{\natexlab{b}})\citenamefont {Zhang}, \citenamefont {Gill},
  \citenamefont {Isenhower}, \citenamefont {Walker},\ and\ \citenamefont
  {Saffman}}]{XZhang2012}%
  \BibitemOpen
  \bibfield  {author} {\bibinfo {author} {\bibfnamefont {X.~L.}\ \bibnamefont
  {Zhang}}, \bibinfo {author} {\bibfnamefont {A.~T.}\ \bibnamefont {Gill}},
  \bibinfo {author} {\bibfnamefont {L.}~\bibnamefont {Isenhower}}, \bibinfo
  {author} {\bibfnamefont {T.~G.}\ \bibnamefont {Walker}}, \ and\ \bibinfo
  {author} {\bibfnamefont {M.}~\bibnamefont {Saffman}},\ }\href@noop {}
  {\bibfield  {journal} {\bibinfo  {journal} {Phys. Rev. A}\ }\textbf {\bibinfo
  {volume} {85}},\ \bibinfo {pages} {042310} (\bibinfo {year}
  {2012}{\natexlab{b}})}\BibitemShut {NoStop}%
\bibitem [{\citenamefont {Nielsen}\ and\ \citenamefont
  {Chuang}(2000)}]{Nielsen2000}%
  \BibitemOpen
  \bibfield  {author} {\bibinfo {author} {\bibfnamefont {M.~A.}\ \bibnamefont
  {Nielsen}}\ and\ \bibinfo {author} {\bibfnamefont {I.~L.}\ \bibnamefont
  {Chuang}},\ }\href@noop {} {\emph {\bibinfo {title} {Quantum computation and
  quantum information}}}\ (\bibinfo  {publisher} {Cambridge University Press,
  Cambridge},\ \bibinfo {year} {2000})\BibitemShut {NoStop}%
\bibitem [{\citenamefont {Jaksch}\ \emph {et~al.}(2000)\citenamefont {Jaksch},
  \citenamefont {Cirac}, \citenamefont {Zoller}, \citenamefont {Rolston},
  \citenamefont {C\^ot\'e},\ and\ \citenamefont {Lukin}}]{Jaksch2000}%
  \BibitemOpen
  \bibfield  {author} {\bibinfo {author} {\bibfnamefont {D.}~\bibnamefont
  {Jaksch}}, \bibinfo {author} {\bibfnamefont {J.~I.}\ \bibnamefont {Cirac}},
  \bibinfo {author} {\bibfnamefont {P.}~\bibnamefont {Zoller}}, \bibinfo
  {author} {\bibfnamefont {S.~L.}\ \bibnamefont {Rolston}}, \bibinfo {author}
  {\bibfnamefont {R.}~\bibnamefont {C\^ot\'e}}, \ and\ \bibinfo {author}
  {\bibfnamefont {M.~D.}\ \bibnamefont {Lukin}},\ }\href@noop {} {\bibfield
  {journal} {\bibinfo  {journal} {Phys. Rev. Lett.}\ }\textbf {\bibinfo
  {volume} {85}},\ \bibinfo {pages} {2208} (\bibinfo {year}
  {2000})}\BibitemShut {NoStop}%
\bibitem [{\citenamefont {Isenhower}\ \emph {et~al.}(2010)\citenamefont
  {Isenhower}, \citenamefont {Urban}, \citenamefont {Zhang}, \citenamefont
  {Gill}, \citenamefont {Henage}, \citenamefont {Johnson}, \citenamefont
  {Walker},\ and\ \citenamefont {Saffman}}]{Isenhower2010}%
  \BibitemOpen
  \bibfield  {author} {\bibinfo {author} {\bibfnamefont {L.}~\bibnamefont
  {Isenhower}}, \bibinfo {author} {\bibfnamefont {E.}~\bibnamefont {Urban}},
  \bibinfo {author} {\bibfnamefont {X.~L.}\ \bibnamefont {Zhang}}, \bibinfo
  {author} {\bibfnamefont {A.~T.}\ \bibnamefont {Gill}}, \bibinfo {author}
  {\bibfnamefont {T.}~\bibnamefont {Henage}}, \bibinfo {author} {\bibfnamefont
  {T.~A.}\ \bibnamefont {Johnson}}, \bibinfo {author} {\bibfnamefont {T.~G.}\
  \bibnamefont {Walker}}, \ and\ \bibinfo {author} {\bibfnamefont
  {M.}~\bibnamefont {Saffman}},\ }\href@noop {} {\bibfield  {journal} {\bibinfo
   {journal} {Phys. Rev. Lett.}\ }\textbf {\bibinfo {volume} {104}},\ \bibinfo
  {pages} {010503} (\bibinfo {year} {2010})}\BibitemShut {NoStop}%
\end{thebibliography}

\begin{thebibliography}{33}%
\makeatletter
\providecommand \@ifxundefined [1]{%
 \@ifx{#1\undefined}
}%
\providecommand \@ifnum [1]{%
 \ifnum #1\expandafter \@firstoftwo
 \else \expandafter \@secondoftwo
 \fi
}%
\providecommand \@ifx [1]{%
 \ifx #1\expandafter \@firstoftwo
 \else \expandafter \@secondoftwo
 \fi
}%
\providecommand \natexlab [1]{#1}%
\providecommand \enquote  [1]{``#1''}%
\providecommand \bibnamefont  [1]{#1}%
\providecommand \bibfnamefont [1]{#1}%
\providecommand \citenamefont [1]{#1}%
\providecommand \href@noop [0]{\@secondoftwo}%
\providecommand \href [0]{\begingroup \@sanitize@url \@href}%
\providecommand \@href[1]{\@@startlink{#1}\@@href}%
\providecommand \@@href[1]{\endgroup#1\@@endlink}%
\providecommand \@sanitize@url [0]{\catcode `\\12\catcode `\$12\catcode
  `\&12\catcode `\#12\catcode `\^12\catcode `\_12\catcode `\%12\relax}%
\providecommand \@@startlink[1]{}%
\providecommand \@@endlink[0]{}%
\providecommand \url  [0]{\begingroup\@sanitize@url \@url }%
\providecommand \@url [1]{\endgroup\@href {#1}{\urlprefix }}%
\providecommand \urlprefix  [0]{URL }%
\providecommand \Eprint [0]{\href }%
\providecommand \doibase [0]{http://dx.doi.org/}%
\providecommand \selectlanguage [0]{\@gobble}%
\providecommand \bibinfo  [0]{\@secondoftwo}%
\providecommand \bibfield  [0]{\@secondoftwo}%
\providecommand \translation [1]{[#1]}%
\providecommand \BibitemOpen [0]{}%
\providecommand \bibitemStop [0]{}%
\providecommand \bibitemNoStop [0]{.\EOS\space}%
\providecommand \EOS [0]{\spacefactor3000\relax}%
\providecommand \BibitemShut  [1]{\csname bibitem#1\endcsname}%
\let\auto@bib@innerbib\@empty

\bibitem [{\citenamefont {Berman}\ and\ \citenamefont {Malinovsky}(2002)}]{BermanMalinovsky}%
  \BibitemOpen
  \bibfield  {author} {\bibinfo {author} {\bibfnamefont {P.~R.}~\bibnamefont
  {Berman}}, \ and\
  \bibinfo {author} {\bibfnamefont {V.~S}~\bibnamefont {Malinovsky}},\ }\href@noop
  {} {\bibfield  {journal} {\bibinfo  {journal} {Principles of Laser Spectroscopy and Quantum Optics, Princeton University Press}\ }\textbf
  {\bibinfo {volume} {}},\ \bibinfo {pages} {} (\bibinfo {year}
  {2011})}\BibitemShut {NoStop}%
\bibitem [{\citenamefont {Petrosyan}\ and\ \citenamefont {Molmer}(2002)}]{Petrosyan2013}%
  \BibitemOpen
  \bibfield  {author} {\bibinfo {author} {\bibfnamefont {D.}~\bibnamefont
  {Petrosyan}}, \ and\
  \bibinfo {author} {\bibfnamefont {K.}~\bibnamefont {M\o{}lmer}},\ }\href@noop
  {} {\bibfield  {journal} {\bibinfo  {journal} {arXiv:1302.0682}\ }\textbf
  {\bibinfo {volume} {}},\ \bibinfo {pages} {} (\bibinfo {year}
  {2013})}\BibitemShut {NoStop}%
\bibitem [{\citenamefont {Beterov}\ \emph {et~al.}(2011)\citenamefont
  {Beterov}, \citenamefont {Tretyakov}, \citenamefont {Entin}, \citenamefont
  {Yakshina}, \citenamefont {Ryabtsev}, \citenamefont {MacCormick},\ and\
  \citenamefont {Bergamini}}]{Beterov2011}%
  \BibitemOpen
  \bibfield  {author} {\bibinfo {author} {\bibfnamefont {I.~I.}\ \bibnamefont
  {Beterov}}, \bibinfo {author} {\bibfnamefont {D.~B.}\ \bibnamefont
  {Tretyakov}}, \bibinfo {author} {\bibfnamefont {V.~M.}\ \bibnamefont
  {Entin}}, \bibinfo {author} {\bibfnamefont {E.~A.}\ \bibnamefont {Yakshina}},
  \bibinfo {author} {\bibfnamefont {I.~I.}\ \bibnamefont {Ryabtsev}}, \bibinfo
  {author} {\bibfnamefont {C.}~\bibnamefont {MacCormick}}, \ and\ \bibinfo
  {author} {\bibfnamefont {S.}~\bibnamefont {Bergamini}},\ }\href@noop {}
  {\bibfield  {journal} {\bibinfo  {journal} {Phys. Rev. A}\ }\textbf {\bibinfo
  {volume} {84}},\ \bibinfo {pages} {023413} (\bibinfo {year}
  {2011})}\BibitemShut {NoStop}%
\bibitem [{\citenamefont {Zhang}\ \emph
  {et~al.}(2012{\natexlab{b}})\citenamefont {Zhang}, \citenamefont {Gill},
  \citenamefont {Isenhower}, \citenamefont {Walker},\ and\ \citenamefont
  {Saffman}}]{XZhang2012}%
  \BibitemOpen
  \bibfield  {author} {\bibinfo {author} {\bibfnamefont {X.~L.}\ \bibnamefont
  {Zhang}}, \bibinfo {author} {\bibfnamefont {A.~T.}\ \bibnamefont {Gill}},
  \bibinfo {author} {\bibfnamefont {L.}~\bibnamefont {Isenhower}}, \bibinfo
  {author} {\bibfnamefont {T.~G.}\ \bibnamefont {Walker}}, \ and\ \bibinfo
  {author} {\bibfnamefont {M.}~\bibnamefont {Saffman}},\ }\href@noop {}
  {\bibfield  {journal} {\bibinfo  {journal} {Phys. Rev. A}\ }\textbf {\bibinfo
  {volume} {85}},\ \bibinfo {pages} {042310} (\bibinfo {year}
  {2012}{\natexlab{b}})}\BibitemShut {NoStop}%
\bibitem [{\citenamefont {Walker}\ and\ \citenamefont
  {Saffman}(2008)}]{Walker2008}%
  \BibitemOpen
  \bibfield  {author} {\bibinfo {author} {\bibfnamefont {T.~G.}\ \bibnamefont
  {Walker}}\ and\ \bibinfo {author} {\bibfnamefont {M.}~\bibnamefont
  {Saffman}},\ }\href@noop {} {\bibfield  {journal} {\bibinfo  {journal} {Phys.
  Rev. A}\ }\textbf {\bibinfo {volume} {77}},\ \bibinfo {pages} {032723}
  (\bibinfo {year} {2008})}\BibitemShut {NoStop}%
	\bibitem [{\citenamefont {Ryabtsev}\ \emph {et~al.}(2010)\citenamefont
  {Ryabtsev}, \citenamefont {Beterov},\citenamefont {Tretyakov}, \citenamefont {Entin},\ and\ \citenamefont
  {Yakshina}}]{Ryabtsev2010}%
  \BibitemOpen
  \bibfield  {author} {\bibinfo {author} {\bibfnamefont {I.~I.}\ \bibnamefont
  {Ryabtsev}}, \bibinfo {author} {\bibfnamefont {D.~B.}\ \bibnamefont
  {Tretyakov}}, \bibinfo {author} {\bibfnamefont {I.~I.}\ \bibnamefont
  {Beterov}}, \bibinfo {author} {\bibfnamefont {V.~M.}\ \bibnamefont {Entin}},
  \bibinfo {author} {\bibfnamefont {E.~A.}\ \bibnamefont {Yakshina}},\ }\href@noop {}
  {\bibfield  {journal} {\bibinfo  {journal} {Phys. Rev. A}\ }\textbf {\bibinfo
  {volume} {82}},\ \bibinfo {pages} {053409} (\bibinfo {year}
  {2010})}\BibitemShut {NoStop}%
\bibitem [{\citenamefont {Brion}\ \emph {et~al.}(2008)\citenamefont {Brion},
  \citenamefont {Pedersen}, \citenamefont {Saffman},\ and\ \citenamefont
  {M\o{}lmer}}]{Brion2008}%
  \BibitemOpen
  \bibfield  {author} {\bibinfo {author} {\bibfnamefont {E.}~\bibnamefont
  {Brion}}, \bibinfo {author} {\bibfnamefont {L.~H.}\ \bibnamefont {Pedersen}},
  \bibinfo {author} {\bibfnamefont {M.}~\bibnamefont {Saffman}}, \ and\
  \bibinfo {author} {\bibfnamefont {K.}~\bibnamefont {M\o{}lmer}},\ }\href@noop
  {} {\bibfield  {journal} {\bibinfo  {journal} {Phys. Rev. Lett.}\ }\textbf
  {\bibinfo {volume} {100}},\ \bibinfo {pages} {110506} (\bibinfo {year}
  {2008})}\BibitemShut {NoStop}%
\bibitem [{\citenamefont {Leung}\ \emph {et~al.}(2011)\citenamefont {Leung},
  \citenamefont {Tauschinsky}, \citenamefont {van~Druten},\ and\ \citenamefont
  {Spreeuw}}]{Spreeuw2011}%
  \BibitemOpen
  \bibfield  {author} {\bibinfo {author} {\bibfnamefont {V.~Y.~F.}~\bibnamefont
  {Leung}}, \bibinfo {author} {\bibfnamefont {A.~S.}\ \bibnamefont {Tauschinsky}},
  \bibinfo {author} {\bibfnamefont {N.~J.}~\bibnamefont {van Druten}}, \ and\
  \bibinfo {author} {\bibfnamefont {R.~J.~C.}~\bibnamefont {Spreeuw}},\ }\href@noop
  {} {\bibfield  {journal} {\bibinfo  {journal} {Quant.~Inf.~Proc}\ }\textbf
  {\bibinfo {volume} {10}},\ \bibinfo {pages} {955} (\bibinfo {year}
  {2011})}\BibitemShut {NoStop}%
\end{thebibliography}

%
\newpage
.
\newpage

{\bf Supplemental Information}

\section{Quantum gate Protocols}

\textit{Single-qubit rotations:}

We define the ensemble states as:
\bea
\ket{\bar 0} &=&\ket{000...000}\\\nonumber
\ket{\bar 1}' &=& \frac{1}{\sqrt{N}}\sum_{j=1}^N \ket{000 ... 1_j ...000}\\\nonumber
\ket{\bar r_0}' &=& \frac{1}{\sqrt{N}}\sum_{j=1}^N \ket{000 ... (r_0)_j ...000}\\\nonumber
\ket{\bar r_1}' &=& \frac{1}{\sqrt{N}}\sum_{j=1}^N \ket{000 ... (r_1)_j ...000}.
\eea

The logical states are $ \ket{\bar 0}$ and $\ket{\bar 1} = e^{\imath\chi_N} \ket{\bar 1}'$. The auxiliary Rydberg states are defined as 
\bea
\ket{\bar r_0} &=&e^{\imath\chi_N}\ket{\bar r_0}' \\\nonumber
\ket{\bar r_1} &=& e^{\imath\chi_N}\ket{\bar r_1}'.
\eea 

Starting with a qubit state $\ket{\psi}=a\ket{\bar 0} + b \ket{\bar 1}$
we perform a sequence of pulses 1-5, shown in Fig.~3(a) of our Letter, giving 
 the sequence of states 
\bea
\ket{\psi_1}&=&a\ket{\bar 0} + ib \ket{\bar r_1}\nonumber\\
\ket{\psi_2}&=&a \ket{\bar r_0} + ib \ket{\bar r_1}\nonumber\\
\ket{\psi_3}&=&a' \ket{\bar r_0} - ib' \ket{\bar r_1}\label{eq.1qubit}\\
\ket{\psi_4}&=&a' \ket{\bar 0} - ib' \ket{\bar r_1}\nonumber\\
\ket{\psi_5}&=&a' \ket{\bar 0} +b' \ket{\bar 1}.\nonumber
\eea

\noindent
The final state $\ket{\psi}=a'\ket{\bar 0} + b' \ket{\bar 1}$ is arbitrary and is selected by the rotation $R(\theta,\phi)$,  in step 3:
 $\begin{pmatrix}a'\\-b'\end{pmatrix}={\bf R}(\theta,\phi)\begin{pmatrix}a\\b\end{pmatrix}$. 

\textit{CNOT:}
Starting with an arbitrary two-qubit state
$\ket{\psi}=a\ket{\bar 0\bar 0} + b \ket{\bar 0 \bar 1} + c\ket{ \bar 1 \bar 0} + d \ket{\bar 1 \bar 1}$
we generate the sequence of states
\bea
\ket{\psi_1}&=&a \ket{{\bar r_0}{\bar 0}} + b \ket{{\bar r_0} {\bar 1}} + c\ket{ {\bar 1}{ \bar 0}} + d \ket{{\bar1} {\bar1}}\nonumber\\
\ket{\psi_2}&=&a \ket{{\bar r_0}{\bar 0}} + b \ket{{\bar r_0} {\bar 1}} + c\ket{ {\bar 1}{ \bar 0}} + id \ket{{\bar1} {\bar r_1}}\nonumber\\
\ket{\psi_3}&=&a \ket{{\bar r_0}{\bar 0}} + b \ket{{\bar r_0} {\bar 1}} + c\ket{ {\bar 1}{ \bar r_0}} + id \ket{{\bar1} {\bar r_1}}\nonumber\\
\ket{\psi_4}&=&ia \ket{{\bar r_1}{\bar 0}} + ib \ket{{\bar r_1} {\bar 1}} + ic\ket{ {\bar 1}{ \bar r_1}} - d \ket{{\bar1} {\bar r_0}}\label{eq.2qubit}\\
\ket{\psi_5}&=&ia \ket{{\bar r_1}{\bar 0}} + ib \ket{{\bar r_1} {\bar 1}} + ic\ket{ {\bar 1}{ \bar r_1}} - d \ket{{\bar1} {\bar 0}}
\nonumber\\
\ket{\psi_6}&=&ia \ket{{\bar r_1}{\bar 0}} + ib \ket{{\bar r_1} {\bar 1}} -c\ket{ {\bar 1}{ \bar 1}} - d \ket{{\bar1} {\bar 0}}
\nonumber\\
\ket{\psi_7}&=&ia \ket{{\bar 0}{\bar 0}} + ib \ket{{\bar 0} {\bar 1}} -c\ket{ {\bar 1}{ \bar 1}} - d \ket{{\bar1} {\bar 0}}
\nonumber.
\eea

\noindent The gate matrix is therefore 
\be
U_{\rm CNOT} = \begin{pmatrix}
i&0&0&0\\
0&i&0&0\\
0&0&0&-1\\
0&0&-1&0\\
\end{pmatrix}\nonumber
\ee
\noindent which can be converted into a standard CNOT gate with a single qubit rotation. 

\section{The numerical model}

We study the population and phase dynamics of the collective states of the 
atomic ensemble which consists of \textit{N} atoms. The vector of the 
probability amplitudes is denoted as $\mathbf{a}\left( {t} \right)$. We  solve the Schr\"odinger 
equation for the vector $\mathbf{a}$, neglecting spontaneous emission~\cite{BermanMalinovsky}:

\be
\label{eq1}
i\hbar\dot{\mathbf{a}}\left({t} \right) = \mathbf{H} \mathbf{a}\left( {t} \right)
\ee

\noindent
\textit{STIRAP:} The atoms in the model have three levels $\ket {0},\,\ket{e},\,\ket{r} $. For 
double STIRAP sequence the matrix of the Hamiltonian is constructed from the 
matrix elements of the operator of dipole interaction of the atomic ensemble 
with two laser fields $\Omega _{1} \left( {t} \right), \, \Omega _{2} \left( {t} \right)$.
Full dipole blockade is represented by complete removal of the amplitudes of 
the collective states with more than one Rydberg excitation $a_{rr} $ from the 
equations. 

As an example, we consider the ensemble which consists of two atoms. The vector of the amplitudes is written as
\be
\mathbf{a} = \left( {{\begin{array}{*{20}c}
 {a_{00}}  \hfill \\
 {a_{0e}}  \hfill \\
 {a_{0r}}  \hfill \\
 {a_{e0}}  \hfill \\
 {a_{ee}}  \hfill \\
 {a_{er}}  \hfill \\
 {a_{r0}}  \hfill \\
 {a_{re}}  \hfill \\
\end{array}} } \right)
\ee

\noindent In the rotating wave approximation for exact two-photon resonance 
the Hamiltonian is written as

\be
\label{eq3}
\mathbf{H} = \frac{{\hbar}}{{2}}\left( {{\begin{array}{*{20}c}
 {0} \hfill & {\Omega _{1}}  \hfill & {0} \hfill & { \Omega _{1} 
} \hfill & {0} \hfill & {0} \hfill & {0} \hfill & {0} \hfill \\
 { \Omega _{1}}  \hfill & { - 2\delta}  \hfill & {\Omega _{2}}  
\hfill & {0} \hfill & {\Omega _{1}}  \hfill & {0} \hfill & {0} \hfill 
& {0} \hfill \\
 {0} \hfill & { \Omega _{2}}  \hfill & {0} \hfill & {0} \hfill & {0} 
\hfill & { \Omega _{1}}  \hfill & {0} \hfill & {0} \hfill \\
 { \Omega _{1}}  \hfill & {0} \hfill & {0} \hfill & { - 2\delta}  
\hfill & { \Omega _{1}}  \hfill & {0} \hfill & {\Omega _{2}}  
\hfill & {0} \hfill \\
 {0} \hfill & { \Omega _{1}}  \hfill & {0} \hfill & { \Omega _{1} 
} \hfill & { - 4\delta}  \hfill & {\Omega _{2}}  \hfill & {0} \hfill & 
{ \Omega _{2}}  \hfill \\
 {0} \hfill & {0} \hfill & { \Omega _{1}}  \hfill & {0} \hfill & { 
\Omega _{2}}  \hfill & { - 2\delta}  \hfill & {0} \hfill & {0} \hfill \\
 {0} \hfill & {0} \hfill & {0} \hfill & { \Omega _{2}}  \hfill & {0} 
\hfill & {0} \hfill & {0} \hfill & { \Omega _{1}}  \hfill \\
 {0} \hfill & {0} \hfill & {0} \hfill & {0} \hfill & {\Omega _{2}}  
\hfill & {0} \hfill & {\Omega _{1}}  \hfill & {-2\delta} \hfill \\
\end{array}} } \right)\quad 
\ee

\noindent The system of equations for the probability amplitudes:

\be
\label{eq4}
\left\{ {\begin{array}{l}
 {i\dot {a}_{00} = \left( {\Omega _{1} /2} \right)a_{0e} + 
\left( {\Omega _{1} /2} \right)a_{e0}}  \\ 
 i\dot {a}_{e0} = - \delta a_{e0} + \left( {\Omega _{1} /2} \right)a_{00} + \left( {\Omega _{1} /2}\right)a_{ee} + \\ \quad
+\left( {\Omega _{2} /2} \right)a_{r0}  \\ 
 i\dot {a}_{0e} = - \delta a_{0e} + \left( {\Omega _{1} /2} \right)a_{00} + \left( {\Omega _{1} /2} 
\right)a_{ee} +\\\quad
+ \left( {\Omega _{2} /2} \right)a_{0r}  \\ 
 {i\dot {a}_{0r} = \left( {\Omega _{1} /2} \right)a_{er} + 
\left( {\Omega _{2} /2} \right)a_{0e}}  \\ 
 {i\dot {a}_{r0} = \left( {\Omega _{1} /2} \right)a_{re} + 
\left( {\Omega _{2} /2} \right)a_{e0}}  \\ 
 i\dot {a}_{ee} = - 2\delta a_{ee} + \left( {\Omega _{1} /2} \right)a_{0e} + \left( {\Omega _{1} /2} 
\right)a_{e0} +\\ \quad
+ \left( {\Omega _{2} /2} \right)a_{er} + 
\left( {\Omega _{2} /2} \right)a_{re} \\ 
 {i\dot {a}_{re} = - \delta a_{re} + \left( {\Omega _{1}/2} \right)a_{r0} + \left( {\Omega _{2}/2 } 
\right)a_{ee}}  \\ 
 {i\dot {a}_{er} = - \delta a_{er} + \left( {\Omega _{1}/2 } \right)a_{0r} + \left( {\Omega _{2}/2 } 
\right)a_{ee}}  \\ 
 \end{array}} \right.
\ee

\noindent \textit{Adiabatic rapid passage:} The atoms in the model have two levels 
$\ket{0},\,\ket{r}$. In the case of perfect
blockade the system of equations describing double sequence of ARP pulses 
for two atoms is written as:

\be
\label{eq5}
\left\{ {\begin{array}{l}
 {i\dot {a}_{00} = \left( {\Omega /2} \right)a_{0r} + \left( {\Omega /2} 
\right)a_{r0}}  \\ 
 {i\dot {a}_{r0} = - \delta \left( {t} \right)a_{r0} + \left( {\Omega /2} 
\right)a_{00}}  \\ 
 {i\dot {a}_{0r} = - \delta \left( {t} \right)a_{0r} + \left( {\Omega /2} 
\right)a_{00}}  \\ 
  \end{array}} \right.
\ee

\noindent We numerically calculate $\left| {a_{00} \left( {t} \right)} \right|^{2}$ 
and $arg\left| {a_{00} \left( {t} \right)} \right|$ to show that population 
inversion is complete and geometric phase is compensated. 

\textit{Account for finite lifetimes}. The finite lifetimes of the 
intermediate state $\ket{e}$ and Rydberg state $\ket{r}$ are taken into account using the density-matrix 
approach, which was applied to an ensemble of interacting atoms. We have 
considered transition $\ket{e} \to \ket{0}$ as closed (the spontaneous decay from the state $\ket{e}$ populates the state $\ket{0}$ only) 
and transition $\ket{r} \to \ket{e}$ 
as open (the spontaneous decay from the state $\ket{r}$ does not populate the states $\ket{e},\ket{0}$)

We have solved the master equation:

\begin{equation}
\label{eq6}
\dot {\rho} \left( {t} \right) = - \frac{{i}}{{\hbar} }\left[ {H,\rho \left( 
{t} \right)} \right] + \hat {L}\rho \left( {t} \right)
\end{equation}

\noindent The Liouvillian superoperator is written as~\cite{Petrosyan2013}

\begin{equation}
\label{eq7}
\hat {L}\rho = \sum\limits_{j} {\left( {\hat {L}_{eg}^{j} \rho + \hat 
{L}_{re}^{j} \rho}  \right)} .
\end{equation}

\noindent Here $\hat {L}_{eg}^{j} \rho = \frac{{1}}{{2}}\gamma _{e} \left[ {2\hat 
{\sigma} _{0e}^{j} \rho \hat {\sigma} _{e0}^{j} - \hat {\sigma} _{ee}^{j} 
\rho - \rho \hat {\sigma} _{ee}^{j}}  \right]$ takes into account decay at 
$\left| {e} \right\rangle \to \left| {0} \right\rangle $ transition for 
j$^{th}$ atom and $\hat {L}_{re}^{j} \rho = \frac{{1}}{{2}}\gamma _{r} 
\left[ { - \hat {\sigma} _{rr}^{j} \rho - \rho \hat {\sigma} _{rr}^{j}}  
\right]$ takes into account decay of the j$^{th}$ atom in the state $\left| 
{r} \right\rangle$ without returning of the population back into the 
system. Here $\hat {\sigma} _{m n} ^{j} = {}_{j}\ket{m}\bra{n}_j$ is a transition operator for 
the j$^{th}$ atom  which undergoes $\ket{m} \to \ket{n}$ transition. We have chosen $\gamma _{e} /\left( {2\pi}  \right) = 5$~MHz 
and $\gamma_{r} /\left( {2\pi}  \right) = 0.8$~kHz. 

The density matrix 
for ensemble containing two atoms is written as $\mathbf{\rho}=\mathbf{aa^{\dag}}$.
The Liouvillian superator for two interacting atoms in the blockade regime 
can be written as:
\begin{widetext}

\be
\begin{array}{l}
 \hat {L}\rho = \frac{{\gamma _{e}} }{{2}} \times \\ 
 \times \left( {{\begin{array}{*{20}c}
 {2\rho _{e0,e0} + 2\rho _{0e,0e}}  \hfill & {2\rho _{e0,ee} - \rho _{00,0e} 
} \hfill & {2\rho _{e0,er}}  \hfill & {2\rho _{0e,ee} - \rho _{00,e0}}  
\hfill & { - 2\rho _{00,ee}}  \hfill & { - \rho _{00,er}}  \hfill & {2\rho 
_{0e,re}}  \hfill & { - \rho _{00,re}}  \hfill \\
 {2\rho _{ee,0e} - \rho _{0e,00}}  \hfill & {2\rho _{ee,ee} - 2\rho _{0e,0e} 
} \hfill & {2\rho _{ee,er} - \rho _{0e,0r}}  \hfill & { - 2\rho _{0e,e0}}  
\hfill & { - 3\rho _{0e,ee}}  \hfill & { - 2\rho _{0e,er}}  \hfill & { - 
\rho _{0e,r0}}  \hfill & { - 2\rho _{0e,re}}  \hfill \\
 {2\rho _{er,e0}}  \hfill & {2\rho _{er,ee} - \rho _{0r,0e}}  \hfill & 
{2\rho _{er,er}}  \hfill & { - \rho _{0r,e0}}  \hfill & { - 2\rho _{0r,ee}}  
\hfill & { - \rho _{0r,er}}  \hfill & {0} \hfill & { - \rho _{0r,re}}  
\hfill \\
 {2\rho _{ee,0e} - \rho _{0e,00}}  \hfill & { - 2\rho _{e0,0e}}  \hfill & { 
- \rho _{e0,0r}}  \hfill & {2\rho _{ee,ee} - \rho _{e0,e0}}  \hfill & { - 
3\rho _{e0,ee}}  \hfill & { - 2\rho _{e0,er}}  \hfill & {2\rho _{ee,re} - 
\rho _{e0,r0}}  \hfill & { - 2\rho _{e0,re}}  \hfill \\
 { - 2\rho _{ee,00}}  \hfill & { - 3\rho _{ee,0e}}  \hfill & { - 2\rho 
_{ee,0r}}  \hfill & { - 3\rho _{ee,e0}}  \hfill & { - 4\rho _{ee,ee}}  
\hfill & { - 3\rho _{ee,er}}  \hfill & { - 2\rho _{ee,r0}}  \hfill & { - 
3\rho _{ee,re}}  \hfill \\
 { - \rho _{er,00}}  \hfill & { - 2\rho _{er,0e}}  \hfill & { - \rho 
_{er,0r}}  \hfill & { - 2\rho _{er,e0}}  \hfill & { - 3\rho _{er,ee}}  
\hfill & { - 2\rho _{er,er}}  \hfill & { - \rho _{er,r0}}  \hfill & { - 
2\rho _{er,re}}  \hfill \\
 {2\rho _{re,0e}}  \hfill & { - \rho _{r0,0e}}  \hfill & {0} \hfill & {2\rho 
_{re,ee} - \rho _{r0,e0}}  \hfill & { - 2\rho _{r0,ee}}  \hfill & { - \rho 
_{r0,er}}  \hfill & {2\rho _{re,re}}  \hfill & { - \rho _{r0,re}}  \hfill \\
 { - \rho _{re,00}}  \hfill & { - 2\rho _{re,0e}}  \hfill & { - \rho 
_{re,0r}}  \hfill & { - 2\rho _{re,e0}}  \hfill & { - 3\rho _{re,ee}}  
\hfill & { - 2\rho _{re,er}}  \hfill & { - \rho _{re,r0}}  \hfill & { - 
2\rho _{re,re}}  \hfill \\
\end{array}} } \right) + \\ \\
 + \frac{{\gamma _{r}} }{{2}}\left( {{\begin{array}{*{20}c}
 {0} \hfill & {0} \hfill & { - \rho _{00,0r}}  \hfill & {0} \hfill & {0} 
\hfill & { - \rho _{00,er}}  \hfill & { - \rho _{00,r0}}  \hfill & { - \rho 
_{00,re}}  \hfill \\
 {0} \hfill & {0} \hfill & { - \rho _{0e,0r}}  \hfill & {0} \hfill & {0} 
\hfill & { - \rho _{0e,er}}  \hfill & { - \rho _{0e,r0}}  \hfill & { - \rho 
_{0e,re}}  \hfill \\
 { - \rho _{0r,00}}  \hfill & { - \rho _{0r,0e}}  \hfill & { - 2\rho 
_{0r,0r}}  \hfill & { - \rho _{0r,e0}}  \hfill & { - \rho _{0r,ee}}  \hfill 
& { - 2\rho _{0r,er}}  \hfill & { - 2\rho _{0r,r0}}  \hfill & { - 2\rho 
_{0r,re}}  \hfill \\
 {0} \hfill & {0} \hfill & { - \rho _{e0,0r}}  \hfill & {0} \hfill & {0} 
\hfill & { - \rho _{e0,er}}  \hfill & { - \rho _{e0,r0}}  \hfill & { - \rho 
_{e0,re}}  \hfill \\
 {0} \hfill & {0} \hfill & { - \rho _{ee,0r}}  \hfill & {0} \hfill & {0} 
\hfill & { - \rho _{ee,er}}  \hfill & { - \rho _{ee,r0}}  \hfill & { - \rho 
_{ee,re}}  \hfill \\
 { - \rho _{er,00}}  \hfill & { - \rho _{er,0e}}  \hfill & { - 2\rho 
_{er,0r}}  \hfill & { - \rho _{er,e0}}  \hfill & { - \rho _{er,ee}}  \hfill 
& { - 2\rho _{er,er}}  \hfill & { - 2\rho _{er,r0}}  \hfill & { - 2\rho 
_{er,re}}  \hfill \\
 { - \rho _{r0,00}}  \hfill & { - \rho _{r0,0e}}  \hfill & { - 2\rho 
_{r0,0r}}  \hfill & { - \rho _{r0,e0}}  \hfill & { - \rho _{r0,ee}}  \hfill 
& { - 2\rho _{r0,er}}  \hfill & { - 2\rho _{r0,r0}}  \hfill & { - 2\rho 
_{r0,re}}  \hfill \\
 { - \rho _{re,00}}  \hfill & { - \rho _{re,0e}}  \hfill & { - 2\rho 
_{re,0r}}  \hfill & { - \rho _{re,e0}}  \hfill & { - \rho _{re,ee}}  \hfill 
& { - 2\rho _{re,er}}  \hfill & { - 2\rho _{re,r0}}  \hfill & { - 2\rho 
_{re,re}}  \hfill \\
\end{array}} } \right)  
 \end{array}
\ee
\end{widetext}

\noindent Decay of the intermediate state leads to breakdown of STIRAP in ensemble of 
interacting atoms if the detuning is not sufficiently large. Decay of the Rydberg state reduces the fidelity of deterministic single-atom excitation.

\textit{STIRAP-based quantum gates:} Validity of the phase compensation for 
the single-atom gate is additionally checked by numerical study of the result 
of the sequence of two single-qubit $\pi /2$ rotations with different phases. 
For this purpose, we included two additional levels $\ket{1},\,\ket{r_1}$ (the state $\ket{r}$ is now denoted as $\ket{r_0}$). The collective states with 
more than one either $r_0 $ or $r_1$ excitation (i.e. $a_{r_0 r_0} ,a_{r_1 r_0} ,a_{r_1 r_1} $) are 
removed from the equations. Below we present the equations used to simulate 
a single-qubit rotation using STIRAP in the ensemble consisting of a single atom:

\be
\label{eq9}
\left\{ {\begin{array}{l}
 {i\dot {a}_{0} = \left( {\Omega _{1} /2} \right)a_{e}}  
\\ 
 {i\dot {a}_{1} = \left( {\Omega _{R} /2} \right)a_{r_1} 
} \\ 
 {i\dot {a}_{e} = - \delta a_{e} + \left( {\Omega_{1} /2} 
\right)a_{0} + \left( {\Omega _{2} /2} \right)a_{r_0}}  \\ 
 {i\dot {a}_{r_0} = \left( {\Omega _{2} /2} \right)a_{e} + 
\left( {\Omega _{G} /2} \right)a_{r_1}}  \\ 
 {i\dot {a}_{r_1} = \left( {\Omega _{R} /2} \right)a_{1} 
+ \left( {\Omega _{G}^{\ast}  /2} \right)a_{r_0}}  \\ 

 \end{array}} \right.
\ee

\noindent 
Here the $\pi $ pulse $\Omega _{R} $ drives $\ket{1}\to\ket{r_1}$ transition and $\Omega _{G}$ 
is an arbitrary Rabi rotation $R\left( {\theta ,\varphi}  \right)$ at 
 $\ket{r_0}\to\ket{r_1}$ transition. Here we take into account that the Rabi frequency $\Omega _{G}$ can be a complex value. The extension to multi-atom ensembles is done in the way similar to Eq.~6. The population of the state $\ket{1}$ is 
calculated at the end of the pulse sequence as function of the phase difference between two $\pi 
/2$ pulses. We have compared it with the outcome of conventional Ramsey-like 
sequence of $\pi /2$ pulses in simple two-level qubit. Similar dynamics was 
observed only if switching of STIRAP detuning is applied.

\section{Experimental implementation}

We now discuss the feasibility of implementation of these gate protocols in a Rydberg blockaded ensemble. To get an isotropic interaction suitable for ensemble blockade we use $ns$ states~\cite{Walker2008}.
The long range interaction strength can be parameterized with a $C_6$ coefficient as $V(n,n')=C_6^{(n,n')}/R^6$ with $R$ the atomic separation.  For Cs $ns$ states the optimum gate fidelity is obtained for $80s$~\cite{XZhang2012}, and the interaction strengths for $|r_0\rangle=|80s_{1/2},m=1/2\rangle, |r_1\rangle=|81s_{1/2},m=1/2\rangle$ are 
$C_6^{(80,80)}=3.2$, $C_6^{(80,81)}=5.1$, $C_6^{(81,81)}=3.7$, in units of $10^6\, {\rm  MHz}\, \mu {\rm m}^6$.  Rydberg $ns$ states can be accessed starting from a ground $s$ state using two-photon STIRAP pulses. These $ns$ states can also be excited with two-photon ARP pulses where  one photon is fixed frequency and one is chirped. Alternatively single photon ARP pulses can be used to access $np$ states. Although the interaction of $np$ states is not isotropic it can be made isotropic in lower dimensional 1- or 2-D lattices by orienting the quantization axis perpendicular to the lattice symmetry plane. For Cs the optimal state 
is  $112p_{3/2}$~\cite{XZhang2012}, and the interaction strengths for $|r_0\rangle=|112p_{3/2},m=3/2\rangle, |r_1\rangle=|113p_{3/2},m=3/2\rangle$ at 90 deg. to the quantization axis are 
$C_6^{(112,112)}=250.$, $C_6^{(112,113)}=820.$, $C_6^{(113,113)}=270.$, in units of $10^6\, {\rm  MHz}\, \mu {\rm m}^6$. We see that for both $ns$ and $np$ states a strong interaction is obtained for all involved  Rydberg states as desired.
The pulse connecting $|r_0\rangle, |r_1\rangle$ can be implemented as a 2-photon electric dipole transition at microwave frequencies via a neighboring opposite parity state. The large transition dipole moments of Rydberg states scaling as $n^2 e a_0$ ($e$ is the electronic charge, $a_0$ is the Bohr radius) render fast microwave pulses straightforward to implement. At $n=80$, a detuning of 1 GHz from the intermediate state, and a very modest $1~\mu\rm W/cm^2$ microwave power level, gives $\sim 25~\rm MHz$ two-photon Rabi frequency.

The  requirements for gates that are insensitive to the value of $N$ are within reach of current experimental capabilities. Optimization of the Rabi frequency and a rigorous determination of the gate fidelity  could be found following the methods of Ref.~\cite{XZhang2012}.  These collective gates require more Rydberg pulses than single atom gates  which will result in different choices for the Rabi frequencies and optimal quantum numbers $n$ compared to the values used for illustration above which are optimal for single atom qubits.
The finite strength of dipole interaction may lead to blockade breakdown. If interaction energy is sufficiently large to prevent multiple Rydberg excitations regardless of the fluctuations of spatial position of the atoms in the ensemble, which could be taken into account using Monte-Carlo simulations~\cite{Ryabtsev2010,Beterov2011}, our method remains feasible.
The collective gates do require that $N$ does not change during the operation of the gate. In any realistic implementation with trapped atoms gate times are significantly  shorter than atom loss times, so this is not an issue. On longer time scales atom loss leads to qubit errors at a physical level, as is also true for a single atom encoding. Ideas for correcting physical errors due to atom loss in ensembles have been presented in Ref.~\cite{Brion2008}. Scalable quantum registers based on atomic ensembles could be created using arrays of optical dipole traps or magnetic traps on magnetic field atom chip~\cite{Spreeuw2011}.

\end{document}